\documentclass[english,conference]{IEEEtran}
\usepackage[T1]{fontenc}
\usepackage[latin9]{inputenc}
\usepackage{amsmath}
\usepackage{amsthm}
\usepackage{amssymb}
\usepackage{graphicx}
\usepackage{epstopdf}

\makeatletter
\theoremstyle{plain}
\newtheorem{thm}{\protect\theoremname}
\theoremstyle{definition}
\newtheorem{defn}[thm]{\protect\definitionname}
\theoremstyle{remark}
\newtheorem{rem}[thm]{\protect\remarkname}
\theoremstyle{plain}
\newtheorem{lem}[thm]{\protect\lemmaname}

\usepackage[latin9]{inputenc}
\usepackage{amsmath}
\usepackage{graphicx}

\IEEEoverridecommandlockouts

\makeatletter

\def\ps@IEEEtitlepagestyle{%
  \def\@oddfoot{\mycopyrightnotice}%
}

\def\mycopyrightnotice{
{\footnotesize  978-1-5090-1138-4/16/\$31.00 �2016 IEEE \hfill 2016 IEEE International Workshop on Information Forensics and Security (WIFS) }
  \gdef\mycopyrightnotice{}
}

\makeatother

\makeatother

\usepackage{babel}
\providecommand{\definitionname}{Definition}
\providecommand{\lemmaname}{Lemma}
\providecommand{\remarkname}{Remark}
\providecommand{\theoremname}{Theorem}

\begin{document}
\title{A Stackelberg Game Perspective on the Conflict Between Machine Learning and Data Obfuscation}



\author{

\IEEEauthorblockN{Jeffrey Pawlick } 
  \IEEEauthorblockA{New York University Tandon School of Engineering \\Department of Electrical and Computer Engineering \\ Email: jpawlick@nyu.edu} 

\and 

\IEEEauthorblockN{Quanyan Zhu} 
  \IEEEauthorblockA{New York University Tandon School of Engineering \\Department of Electrical and Computer Engineering \\ Email: quanyan.zhu@nyu.edu} 
    
\thanks{This work is partially supported by the grant CNS-1544782,  EFRI-1441140  and  SES-1541164  from National Science Foundation.}
}



\maketitle
\begin{abstract}
Data is the new oil; this refrain is repeated extensively in the age
of internet tracking, machine learning, and data analytics. As data
collection becomes more personal and pervasive, however, public pressure
is mounting for privacy protection. In this atmosphere, developers
have created applications to add noise to user attributes visible
to tracking algorithms. This creates a strategic interaction between
trackers and users when incentives to maintain privacy and improve
accuracy are misaligned. In this paper, we conceptualize this conflict
through an $N+1$-player, augmented Stackelberg game. First a machine
learner declares a privacy protection level, and then users respond
by choosing their own perturbation amounts. We use the general frameworks
of differential privacy and empirical risk minimization to quantify
the utility components due to privacy and accuracy, respectively.
In equilibrium, each user perturbs her data independently, which leads
to a high net loss in accuracy. To remedy this scenario, we show that
the learner improves his utility by proactively perturbing the data
himself. While other work in this area has studied privacy markets
and mechanism design for truthful reporting of user information, we
take a different viewpoint by considering both user and learner perturbation.
\end{abstract}

\section{Introduction\label{sec:Intro} }

In the modern digital ecosystem, users leave behind rich trails of
behavioral information. On the internet, websites send user data to
third-party trackers such as advertising agencies, social networking
sites, and data analytic companies \cite{mayer_third-party_2012}.
Tracking is not limited, of course, to the internet. The internet
of things (IoT) is a phenomenon that refers to the standardization
and integration of communications between physical devices in a way
that mimics the connection of computers on the internet. IoT devices
such as smartwatches include accelerometers, heart rate sensors, and
sleep trackers that measure and upload data about users' physical
and medical conditions \cite{swan_sensor_2012}. Data from these applications
data can be used to improve product or service quality or to drive
social change. For example, continuous glucose monitors can provide
closed-loop blood glucose control for users with diabetes \cite{continuousGlucose,parker1999model}.
The smart grid and renewable energy also stand to benefit from developments
in networks of sensors and actuators \cite{baheti2011cyber}.

\subsection{Privacy in Machine Learning}

While these technologies promise positive impacts, they also threaten
privacy. Specifically, the IoT involves new threats in the form of
information access, because devices may directly collect sensitive
information such as health and location data \cite{_internet_2015}.
In addition, the pervasiveness of tracking and the development of
analytics have enabled learners to infer habits and physical conditions
over time. These inferences may run even to the granularity of ``a
user's mood; stress levels; personality type; bipolar disorder; demographics''
\cite{peppet_regulating_2014}. These are unprecedented degrees of
access to user information. This access has prompted both qualitative
and quantitative privacy research.

While several methods have been developed to quantify privacy, we
focus on one particular notion in this paper. Proposed by Cynthia
Dwork, differential privacy is a mathematical framework which gives
probable limits on the disclosure risks that individuals incur by
participating in a database \cite{dwork2006differential,dwork2008difficulties,dwork2014algorithmic}.
Using DP, learning algorithms can publish a guarantee on the amount
of information disclosed: namely, the constant often denoted $\epsilon_{p}.$
Currently, however, there seems to be little incentives for trackers
to adopt DP methods.

\subsection{User Obfuscation Technologies}

To remedy this situation, developers have begun to help users perturb
data on their own. Finn and Nissenbaum describe two examples: \emph{CacheCloak}
and \emph{TrackMeNot} \cite{brunton2015obfuscation}. \emph{TrackMeNot}
is a browser extension that generates decoy search queries in order
to prevent trackers from assembling accurate profiles of its users
\cite{howe_trackmenot:_2009}. In the realm of IoT, \emph{CacheCloak}
provides a way for users to access location-based services without
revealing their exact geographical positions \cite{meyerowitz_hiding_2009}.
The app predicts multiple possibilities for the path of a user, and
then retrieves location-based information for each path. This means
that an adversary tracking the requests is left with many possible
paths rather than a unique one. As another example, the browser extension
\emph{ScareMail} adds words relevant to terrorism to every email that
a user issues, postulating that wide adoption of this technique would
make dragnet surveillance difficult \cite{grosser2014scare}. Apparently,
however, such privacy protection involves costs not only for governments
but also for the whole population of users.

\subsection{Learner-User Interaction}

\begin{figure}
\begin{centering}
\includegraphics[width=0.9\columnwidth]{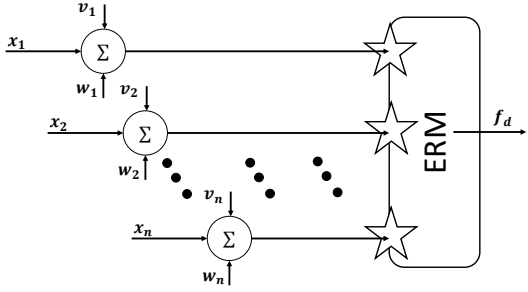}
\par\end{centering}

\caption{\label{fig:dataFlow}Data flow in the obfuscation-tracking model.
Users $1,\ldots,N$ have data $\mathbf{x_{i}}$ with labels $y_{i}.$
Before submitting this data to a classifier, the users add noise $\mathbf{v_{i}}\sim\mathcal{V}_{i},$
and the learner can add noise $\mathbf{w_{i}}\sim\mathcal{W}.$ The
classifier is $\mathbf{f_{d}}.$ The stars indicate that the learner
is the privacy adversary.}
\end{figure}

This conflict can be studied by an interaction between $N$ users
and a machine learner. This data flow in Fig. \ref{fig:dataFlow}.
In general, both the users and the learner could be interested in
the privacy and accuracy of the learning outcome. But these incentives
are probably not aligned. Hence the interaction is strategic, and
aptly studied by game theory. 

We model the user-learner interaction as a two-step process in which
the learner first announces his perturbation level, and then the users
respond by implementing their own perturbation. This is a realistic
assumption, since a critical aspect of DP is the ability to publish
measurable privacy guarantees. Knowing this protection, users can
decide whether to add their own perturbation in order to further protect
their information. The dynamic, two-stage nature of this interaction
suggests the framework of Stackelberg games \cite{von_stackelberg_marktform_1934,basar_stackelberg_1999}.

\subsection{Content and Contributions}

In Section \ref{sec:ERM_DP_Mod} we describe the machine learning
technique of Empirical Risk Minimization (ERM) and the framework of
DP. Then, in Section \ref{sec:interaction}, we employ DP to quantify
utility loss due to privacy compromise, and ERM to quantify utility
gained through an accurate predictor. In Section \ref{sec:solCon},
we review the solution concept of Stackelberg equilibrium, and we
study the equilibrium in Section \ref{sec:ana}. Finally, we discuss
the importance of the results in Section \ref{sec:Conclu}.

In summary, this paper presents the following contributions:
\begin{enumerate}
\item We create a Stackelberg game model to study the conflict between tracking
and obfuscation.
\item Our model uses the framework of ERM to quantify accuracy, and DP to
quantify privacy loss. These frameworks are sufficiently broad to
be used for many different application areas.
\item We find that, while the accuracy levels of all of the users are interdependent,
the strategic optimal perturbation level for each user is independent
of the perturbation levels of all of the other users (Remark \ref{rem:indep}). 
\item In equilibrium, if the learning algorithm adds sufficient perturbation,
it can dissuade the users from obfuscating the data themselves (Remark
\ref{rem:BR}). 
\item When the cost of user perturbation is high, protecting user privacy
by proactively perturbing is incentive-compatible for the learner
(Remark \ref{rem:learner}).
\end{enumerate}

\subsection{Related Work}

In order to address incentive-compatibility, a vein of research has
arisen in \emph{privacy markets}. In \cite{ghosh2013selling}, a learner
computes a sum of the private bits of a set of users and tries to
either maximize accuracy or minimize cost. This paper assumes that
users report their data truthfully but can misrepresent their individual
valuation of their privacy. Later authors interchanged these assumptions
\cite{xiao2013privacy}. In work by Chessa \emph{et al}. \cite{chessa2015short,chessa2015game},
users play a multiple person, prior-commitment game, which determines
how much they perturb. The present paper differs from all four of
these works because it considers the learner as an additional strategic
player. Shokri \emph{et al.} \cite{shokri2012protecting} formulate
a Stackelberg game for preserving location privacy. In this game,
the user is the leader and the learner is the follower. After the
user chooses a perturbation strategy, the learner chooses an optimal
reconstruction of the user's location. By contrast, in our model the
learner chooses a \emph{promised level of privacy protection} before
the user acts, which makes the learner a Stackelberg leader. Lastly,
unlike all of the previous works, our model uses both empirical risk
minimization and differential privacy.

\section{Empirical Risk Minimization and Differential Privacy Models\label{sec:ERM_DP_Mod} }

Consider an interaction between a set of users $i\in S=\left\{ 1,\ldots,N\right\} $
and a learner $L,$ in which users submit possibly-perturbed data
to $L,$ and $L$ releases a statistic or predictor of the data $\mathbf{f_{d}}$
(hereafter, an \emph{output}). Assume that the data generating process
is a random variable $\mathcal{Z}$ with a fixed but unknown distribution.
Denote the realized data by $\mathbf{z_{i}}\overset{\text{i.i.d.}}{\sim}\mathcal{Z},\,i\in S.$
Each data point is composed of a feature vector $\mathbf{x_{i}}\in\mathbb{R}^{d}$
and a label $y_{i}\in\left\{ -1,1\right\} .$ The goal of the learner
$L$ is to predict $y_{i}$ given $\mathbf{x_{i}},$ based on the
trained classifier or predictor $\mathbf{f_{d}}.$ 

In general, privacy loss can occur 1) with respect to $L$, and 2)
with respect to the public who observes the output of the ERM. In
order to narrow the scope of this paper, we consider information disclosure
with respect to $L.$ In addition, information can be leaked through
1) the attributes $\mathbf{x_{i}}$ and 2) the labels $y_{i}.$ We
focus on loss due to $\mathbf{x_{i}}$, although analysis using $y_{i}$
would follow many of the same principles. 

With the threat of user perturbation, we investigate whether it is
advantageous for $L$ to proactively protect the privacy of the users.
Thus, we allow $L$ to perturb the submitted data, also \emph{before}
she views it\footnote{$L$ must use a trusted execution environment in order to perturb
the data. Alternatively, $L$ may accomplish this purpose by collecting
data at a lower granularity from the users.}. Assume that $L$ adds noise with the same variance to each data
point $\mathbf{x_{i}}.$ For $i\in S,$ $k\in1,\ldots,d,$ the learner
draws $w_{i}^{\left(k\right)}\overset{\text{i.i.d}}{\sim}\mathcal{W},$
where $\mathcal{W}$ is a mean-zero Gaussian random variable\footnote{While DP often considers Laplace noise, we use Gaussian noise for
reasons of mathematical convenience. } with standard deviation $\sigma_{L}.$ Then the user adds noise $v_{i}^{\left(k\right)}\overset{\text{i.i.d.}}{\sim}\mathcal{V}_{i},$
$k\in1,\ldots,d,$ where $\mathcal{V}_{i}$ is also Gaussian. The
perturbed data points are given by $\mathbf{\tilde{x}_{i}}=\mathbf{x_{i}}+\mathbf{v_{i}}+\mathbf{w_{i}},$
$i\in S.$ Figure \ref{fig:dataFlow} summarizes this flow of data.

\subsection{Empirical Risk Minimization\label{sub:Empirical-Risk-Minimization}}

In empirical risk minimization, $L$ calculates a value of output
$\mathbf{f_{d}}\in\mathbf{F}$ that minimizes the empirical risk,
\emph{i.e.}, the total penalty due to imperfect classification of
the realized data. Define the \emph{loss function} $l\left(\mathbf{\tilde{z}_{i}},\mathbf{f}\right),$
which expresses the penalty due to a single perturbed data point $\mathbf{z_{i}}$
for the output $\mathbf{f}.$ Next let $\Lambda\geq0$ be a constant
and $R\left(\mathbf{f}\right)$ be a regularization term. For $\mathbf{z_{i}}$
in the database $D,$ the total empirical risk is $J\left(\mathbf{f},D\right)=\Lambda R\left(\mathbf{f}\right)+\frac{1}{N}\underset{i}{\sum}l\left(\mathbf{z_{i}},\mathbf{f}\right).$
$L$ obtains $\mathbf{f_{d}}$ given by Eq. \ref{eq:thetaD}. Unperturbed
data gives the classifier $\mathbf{f^{\dagger}}$ in Eq. \ref{eq:thetaDagger}:
\begin{equation}
\mathbf{f_{d}}=\underset{\mathbf{f}\in\mathbf{F}}{\arg\min}\,\Lambda R\left(\mathbf{f}\right)+\frac{1}{N}\underset{i}{\sum}l\left(\mathbf{\tilde{z}_{i}},\mathbf{f}\right),\label{eq:thetaD}
\end{equation}
\begin{equation}
\mathbf{f^{\dagger}}=\underset{\mathbf{f}\in\mathbf{F}}{\arg\min}\,\Lambda R\left(\mathbf{f}\right)+\frac{1}{N}\underset{i}{\sum}l\left(\mathbf{z_{i}},\mathbf{f}\right).\label{eq:thetaDagger}
\end{equation}

\emph{Expected loss} provides a measure of the accuracy of the output
of ERM. Let $\mathbf{f^{*}}$ denote the $\mathbf{f}$ which minimizes
the \emph{expected} loss for unperturbed data:
\begin{equation}
\mathbf{f^{*}}=\underset{\mathbf{f}\in\mathbf{F}}{\text{argmin}}\,\mathbb{E}\left\{ \Lambda R\left(\mathbf{f}\right)+l\left(\mathcal{Z},\mathbf{f}\right)\right\} .
\end{equation}
This forms a reference to which the expected loss of $\mathbf{f_{d}}$
on data $\mathcal{Z}$ can be compared. Let $\epsilon_{g}$ be a positive
scalar that bounds the difference in expected loss between the perturbed
classifier and the population-optimal classifier. This quantity is
given by
\begin{equation}
\mathbb{E}\left\{ \Lambda R\left(\mathbf{f_{d}}\right)+l\left(\mathcal{Z},\mathbf{f_{d}}\right)\right\} \leq\mathbb{E}\left\{ \Lambda R\left(\mathbf{f^{*}}\right)+l\left(\mathcal{Z},\mathbf{f^{*}}\right)\right\} +\epsilon_{g}.\label{eq:utilDif}
\end{equation}
We use this difference to formulate the accuracy component of utility
in Section \ref{sec:interaction}.

\subsection{Differential Privacy\label{sub:DP}}

Let $\mathcal{A}\left(*\right)$ denote an algorithm and $D$ denote
a database. Let $D'$ denote a database that differs from $D$ by
only one entry (\emph{e.g.}, the entry of the user under consideration).
Let $c$ be some set among all possible sets $C$ in which the output
of the algorithm $\mathcal{A}$ may fall. Then Definition \ref{def:DP}
quantifies privacy using the framework of DP \cite{chaudhuri2011differentially,dwork2006differential}.
\begin{defn}
\label{def:DP}($\epsilon_{p}$-DP) - An algorithm $\mathcal{A}\left(B\right)$
taking values in a set $C$ provides $\left(\epsilon_{p},\delta\right)$-differential
privacy if, for all $D,$ $D'$ that differ in at most one entry,
and for all $c\in C,$
\begin{equation}
\mathbb{P}\left\{ \mathcal{A}\left(D\right)\in c\right\} \leq\exp\left(\epsilon_{p}\right)\mathbb{P}\left\{ \mathcal{A}\left(D'\right)\in c\right\} +\delta.\label{eq:DP}
\end{equation}

\end{defn}
For a fixed $\delta,$ the degree of randomness determines the privacy
level $\epsilon_{p}$. Lower values of $\epsilon_{p}$ correspond
to more privacy. That randomness is attained through the noise added
in the forms of $\mathcal{V}$ and $\mathcal{W}.$

\section{Dynamic User-Learner Interaction\label{sec:interaction}}

We now use the methods for quantification of accuracy and privacy
described in Section \ref{sec:ERM_DP_Mod} as components of utility
functions for the users and the learner.

\subsection{Utility Functions}

Let $U_{S}^{i}\left(\sigma_{L},\sigma_{S}^{-i},\sigma_{S}^{i}\right)$
give the utility that each user $i$ receives when the learner chooses
perturbation $\sigma_{L},$ user $i$ chooses perturbation level $\sigma_{S}^{i},$
and all of the other users choose perturbation levels $\sigma_{S}^{-i}\triangleq\{\sigma_{S}^{j}\}_{j\in S\backslash i}.$
Similarly, let $U_{L}\left(\sigma_{L},\sigma_{S}\right)$ be a utility
function for the learner, $L,$ where $\sigma_{S}\triangleq\{\sigma_{S}^{j}\}_{j\in S}.$
The utility functions have components due to accuracy, privacy, and
cost of perturbation. Note that each user's perturbation affects her
own privacy directly, but affects her accuracy only after ERM based
on all users' data points.

\subsection{Accuracy Component of Utility}

The accuracy component of utility is determined by the accuracy of
$\mathbf{f_{d}}$ as a function of $\sigma_{L}$ and $\sigma_{S}.$
This accuracy is in terms of the difference $\epsilon_{g}$ in expected
loss between the perturbed and unperturbed classifiers (Eq. \ref{eq:utilDif}).
The relationship is summarized by Theorem \ref{thm:accConstant}.
\begin{thm}
\label{thm:accConstant}(Accuracy Constant $\epsilon_{g}$) For a
fixed distribution $\mathcal{Z},$ define expected loss by $\hat{J}\left(\mathbf{f}\right)=\mathbb{E}_{\left(\mathbf{x},y\right)\sim\mathcal{Z}}\left\{ l\left(\mathbf{f}^{T}\mathbf{x},y\right)\right\} +\frac{\Lambda}{2}\left\Vert \mathbf{f}\right\Vert ^{2}.$
Then the dependence of the difference in expected loss on the user
and learner perturbation levels is given, with some chosen probability,
by 
\begin{equation}
\hat{J}\left(\mathbf{f_{d}}\right)-\hat{J}\left(\mathbf{f}^{*}\right)=\propto\frac{1}{n\Lambda^{2}}\left(\sigma_{L}^{2}+\underset{i}{\sum}\frac{1}{n}\left(\sigma_{S}^{i}\right)^{2}\right).\label{eq:accComp}
\end{equation}
\end{thm}
\begin{IEEEproof}
See Appendix.
\end{IEEEproof}
Equation \ref{eq:accComp} will be used to formulate the utility component
of accuracy in Subsection \ref{sub:totalUtil}.

\subsection{Privacy Component of Utility}

The privacy of the data $\mathbf{x_{i}},$ $i\in S$ submitted to
$L$ is achieved by the Gaussian mechanism \cite{dwork2014algorithmic}.
\begin{defn}
(Gaussian Mechanism) Let a database consist of entries $\mathbf{x}\in\mathbb{X},$
and denote the space of all possible databases by $\mathbb{N}^{\left|\mathbb{X}\right|}.$
Let $\mathcal{A}:\,\mathbb{N}^{\mathbb{X}}\to\mathbb{R}^{d}$ be an
arbitrary $d$-dimensional function. The Gaussian Mechanism with parameter
$\sigma$ adds noise with mean $0$ and variance $\sigma^{2}$ to
each of the $d$ components of the output.
\end{defn}
In \cite{dwork2014algorithmic}, Dwork and Roth obtain a differential
privacy guarantee for the Gaussian Mechanism, solved here for $\epsilon_{p}.$
We use the fact that the total perturbation $\mathcal{V}_{i}+\mathcal{W}$
has standard deviation $\sqrt{\sigma_{L}^{2}+\left(\sigma_{S}^{i}\right)^{2}}.$
\begin{thm}
Let $S\left(\mathcal{A}\right)$ denote the $L_{2}$ sensitivity of
$\mathcal{A}.$ For $\epsilon_{p}\in\left(0,1\right),$ the Gaussian
Mechanism achieves $\left(\epsilon_{p},\delta\right)$-differential
privacy if $\sigma$ satisfies
\begin{equation}
\epsilon_{p}=\frac{2\sqrt{2\ln\left(1.25/\delta\right)}}{\sigma}\propto\frac{1}{\sqrt{\sigma_{L}^{2}+\left(\sigma_{S}^{i}\right)^{2}}}.
\end{equation}
 
\end{thm}

\subsection{Perturbation Cost Component of Utility}

How can the cost of perturbation be defined? Currently, many applications
that perturb user data are free. This is true of \emph{TrackMeNot},
\emph{CacheCloak}, and \emph{ScareMail}. On the other hand, users
experience some non-monetary cost (\emph{e.g.}, time, learning curve,
aversion to degrading quality of data). This cost is arguably flat
with respect to perturbation amount. Define the perturbation components
of utility for variances of $\sigma_{L}^{2}$ and $\left(\sigma_{S}^{i}\right)^{2}$
by $\bar{N}_{L}\mathbf{1}_{\left\{ \sigma_{L}>0\right\} }$ and $\bar{N}_{S}^{i}\mathbf{1}_{\left\{ \sigma_{S}^{i}>0\right\} },$
respectively, where $\bar{N}_{L}$ and $\bar{N}_{S}^{i}$ are positive
coefficients.

\subsection{Total Utility Functions\label{sub:totalUtil}}

The utility functions in are given by combining the utility terms
due to accuracy, privacy, and perturbation cost. Define $\bar{G}_{L}$
and $\bar{G}_{S}^{i}$ as positive values of the unperturbed accuracy
to the learner and to each user $i,$ respectively. Let $\gamma_{L}$
and $\gamma_{S}^{i}$ adjust the rate of utility loss due to accuracy.
Next, let $\bar{P}_{S}^{i}$ denote the maximum privacy loss to user
$i,$ which she incurs if the data is not perturbed at all\footnote{We have made the privacy term for $L$ proportional to the average
privacy of the users, based on an assumption that $L$ benefits from
adding value in the form of privacy to the users. Other parameters
are used to set the relative importance of privacy and accuracy for
the users.}. Finally, we use $\rho_{S}^{i}>0$ to scale the rate of privacy loss
for user $i.$ Now the utility functions are given by:\vspace{-0.8cm}

\begin{multline}
U_{L}\left(\sigma_{L},\sigma_{S}\right)=\bar{G}_{L}-\frac{\gamma_{L}}{n\Lambda^{2}}\left(\sigma_{L}^{2}+\underset{i}{\sum}\frac{1}{n}\left(\sigma_{S}^{i}\right)^{2}\right)\\
-\frac{1}{N}\underset{i}{\sum}\frac{\bar{P}_{S}^{i}}{1+\rho_{S}^{i}\sqrt{\sigma_{L}^{2}+\left(\sigma_{S}^{i}\right)^{2}}}-\bar{N}_{L}\mathbf{1}_{\left\{ \sigma_{L}>0\right\} },
\end{multline}
\vspace{-0.4cm}
\begin{multline}
U_{S}^{i}\left(\sigma_{L},\sigma_{S}^{-i},\sigma_{S}^{i}\right)=\bar{G}_{S}^{i}-\frac{\gamma_{S}^{i}}{n\Lambda^{2}}\left(\sigma_{L}^{2}+\underset{i}{\sum}\frac{1}{n}\left(\sigma_{S}^{i}\right)^{2}\right)\\
-\frac{\bar{P}_{S}^{i}}{1+\rho_{S}^{i}\sqrt{\sigma_{L}^{2}+\left(\sigma_{S}^{i}\right)^{2}}}-\bar{N}_{S}^{i}\mathbf{1}_{\left\{ \sigma_{S}^{i}>0\right\} }.
\end{multline}

\subsection{Independence of the Users}

Notice that the derivative of $U_{S}^{i}\left(\sigma_{L},\sigma_{S}^{-i},\sigma_{S}^{i}\right)$
with respect to $\sigma_{S}^{i}$ is not a function of any $\sigma_{S}^{j}$
for $j\in S\backslash i.$ This leads to the following remark.
\begin{rem}
\label{rem:indep}The optimal perturbation level for each user is
independent of the actions of the other users.
\end{rem}
In fact, this is analogous to the prisoner's dilemma, in which the
utilities of the players are coupled although the \emph{optimal actions}
are not. The independence of the users provides the following useful
fact.
\begin{rem}
\label{rem:solveIndep}The equilibrium of the $N+1$-player game can
be found as by considering all of the users as one aggregate player,
since their strategies are independent. The solution concept is a
traditional Stackelberg equilibrium.
\end{rem}

\section{Solution Concept\label{sec:solCon}}

\begin{figure}
\begin{centering}
\includegraphics[width=0.75\columnwidth]{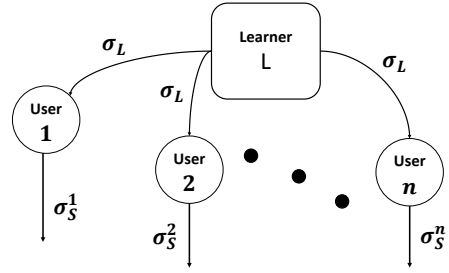}
\par\end{centering}

\caption{\label{fig:Stack}Stackelberg game interaction between the learner
and the set of users. This diagram depicts the flow of actions, rather
than the flow of data.}
\end{figure}

Figure \ref{fig:Stack} depicts the flow of actions in the Stackleberg
game. $L$ chooses perturbation level $\sigma_{L},$ which he announces.
Then the users respond with their own perturbation levels $\sigma_{S}^{i},$
$i\in S.$ The users' strategies are independent of each other, but
$L$ must act in anticipation of the actions of the set of all of
the users.

Definition \ref{def:stack} describes a Stackelberg equilibrium. Define
$BR_{S}^{i}:\,\mathbb{R}_{+}\to\mathbb{R}_{+},$ such that $\sigma_{S}^{i}=BR_{S}^{i}\left(\sigma_{L}\right)$
gives strategy $\sigma_{S}^{i}$ which best responds to the learner's
perturbation level $\sigma_{L},$ and let $BR_{S}\left(\sigma_{L}\right)\triangleq\left\{ BR_{S}^{i}\left(\sigma_{L}\right)\right\} _{i\in S}.$ 
\begin{defn}
\label{def:stack}(Stackelberg Equilibrium) The strategy profile $\left(\sigma_{L},\left\{ \sigma_{S}^{i}\right\} _{i\in S}\right)$
is a Stackelberg equilibrium if, $\forall i\in S,$ 
\begin{equation}
\sigma_{S}^{i*}=BR_{S}^{i}\left(\sigma_{L}^{*}\right)\triangleq\underset{\sigma_{S}^{i}}{\arg\max\:}U_{S}^{i}\left(\sigma_{L}^{*},\sigma_{S}^{-i*}\sigma_{S}^{i}\right),\label{eq:BR}
\end{equation}
 
\begin{equation}
\sigma_{L}^{*}=\underset{\sigma_{L}}{\arg\max\:}U_{L}\left(\sigma_{L},BR_{S}\left(\sigma_{L}\right)\right).\label{eq:sigL_from_BR}
\end{equation}

\end{defn}
The order of solution is the reverse of the chronological order; the
best response function $BR_{S}^{i}\left(\sigma_{L}^{*}\right)$ must
be found first from Eq. \ref{eq:BR}. Then it is possible to solve
Eq. \ref{eq:sigL_from_BR}.

\section{Analysis\label{sec:ana}}

\begin{figure*}
\begin{centering}
\includegraphics[width=0.25\textwidth]{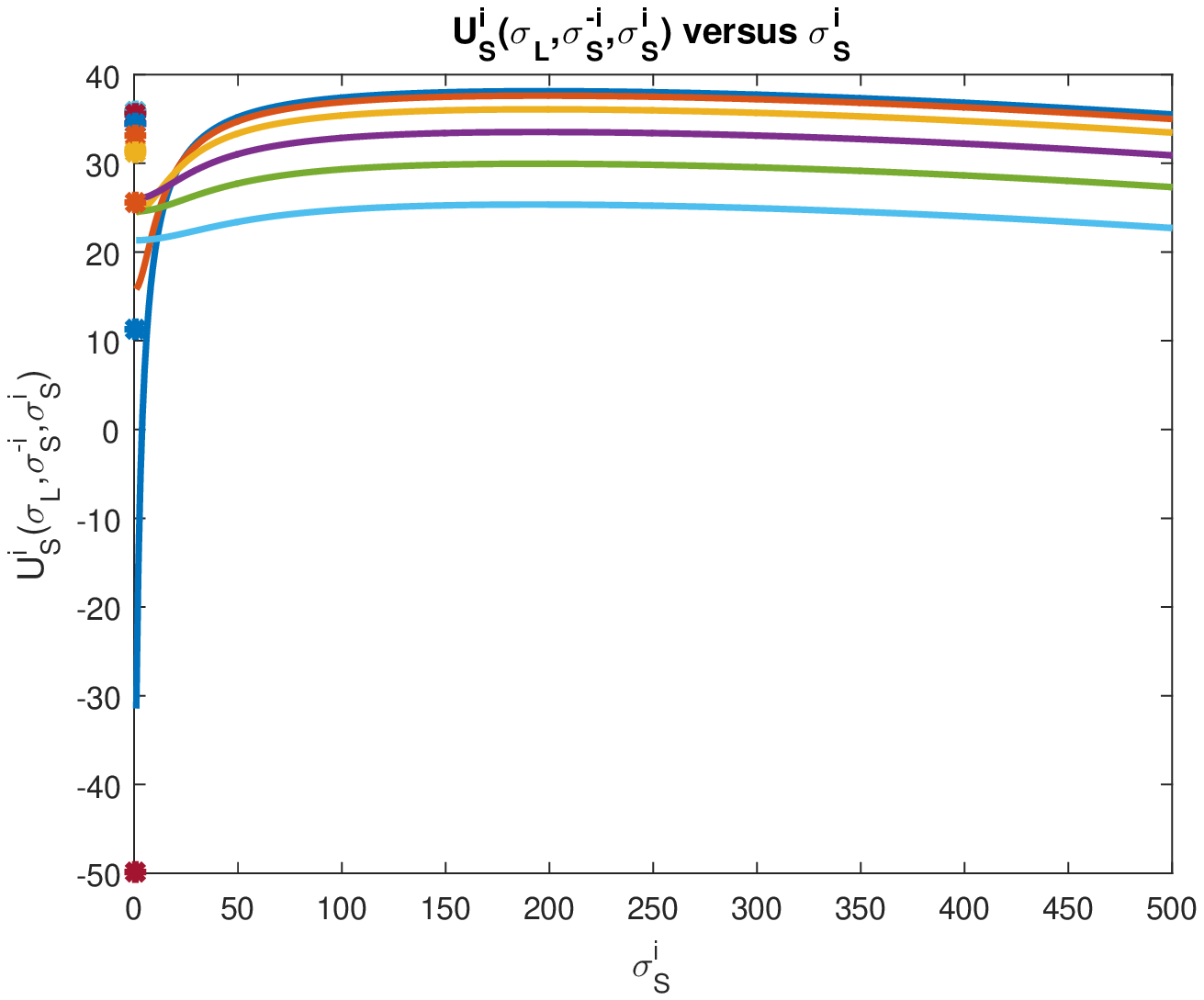}\includegraphics[width=0.25\textwidth]{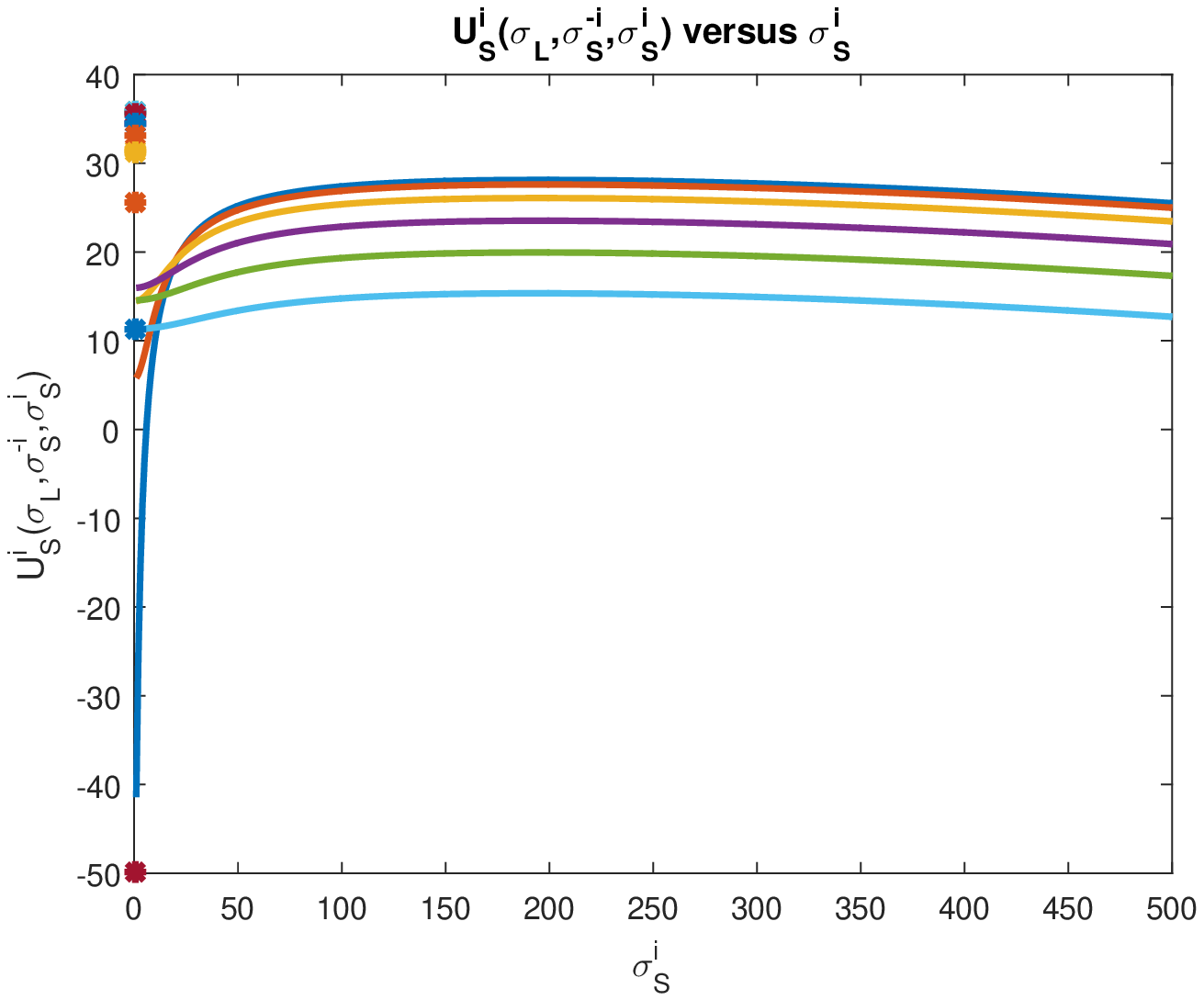}\includegraphics[width=0.25\textwidth]{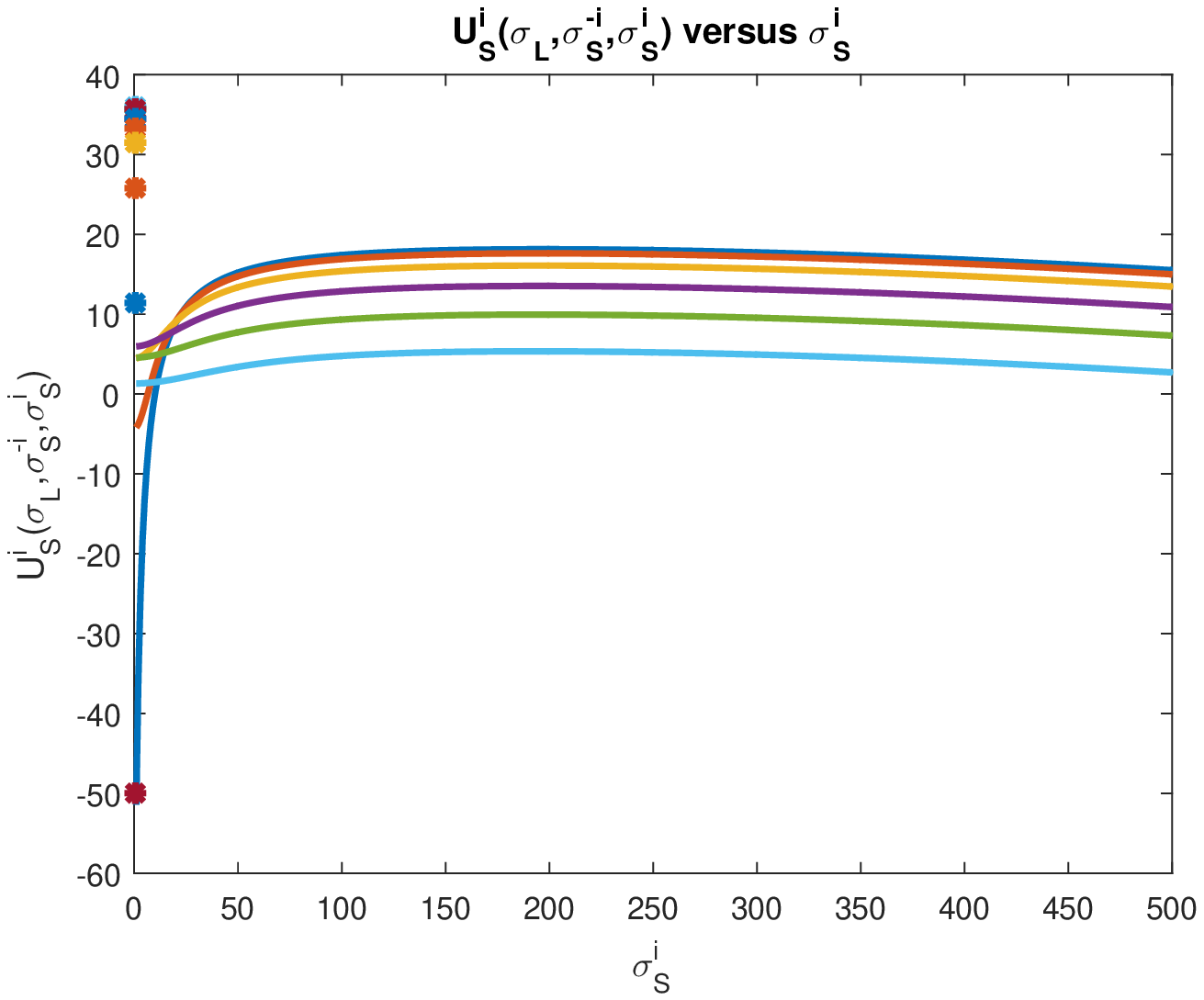}
\par\end{centering}

\begin{centering}
\includegraphics[width=0.25\textwidth]{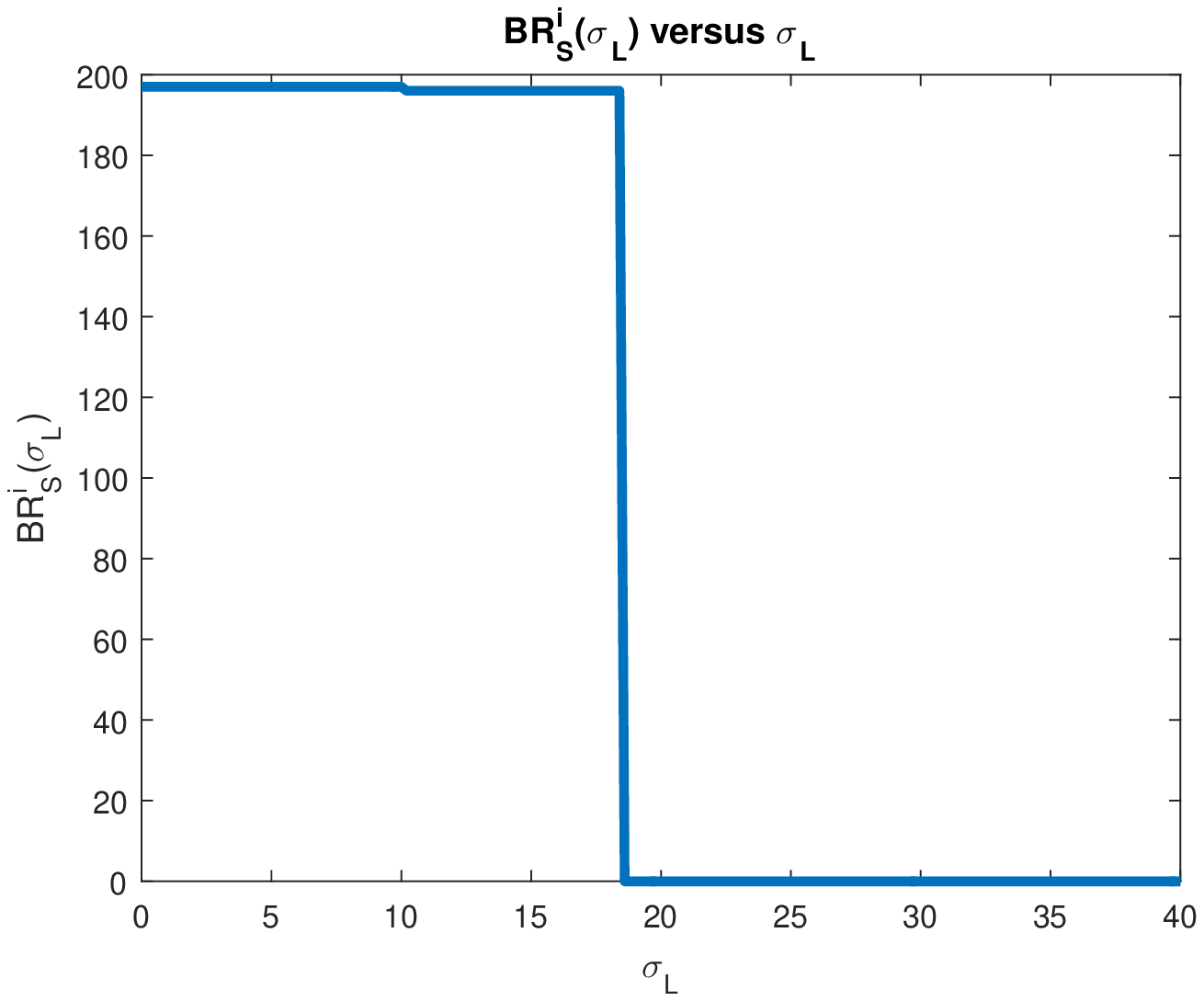}\includegraphics[width=0.25\textwidth]{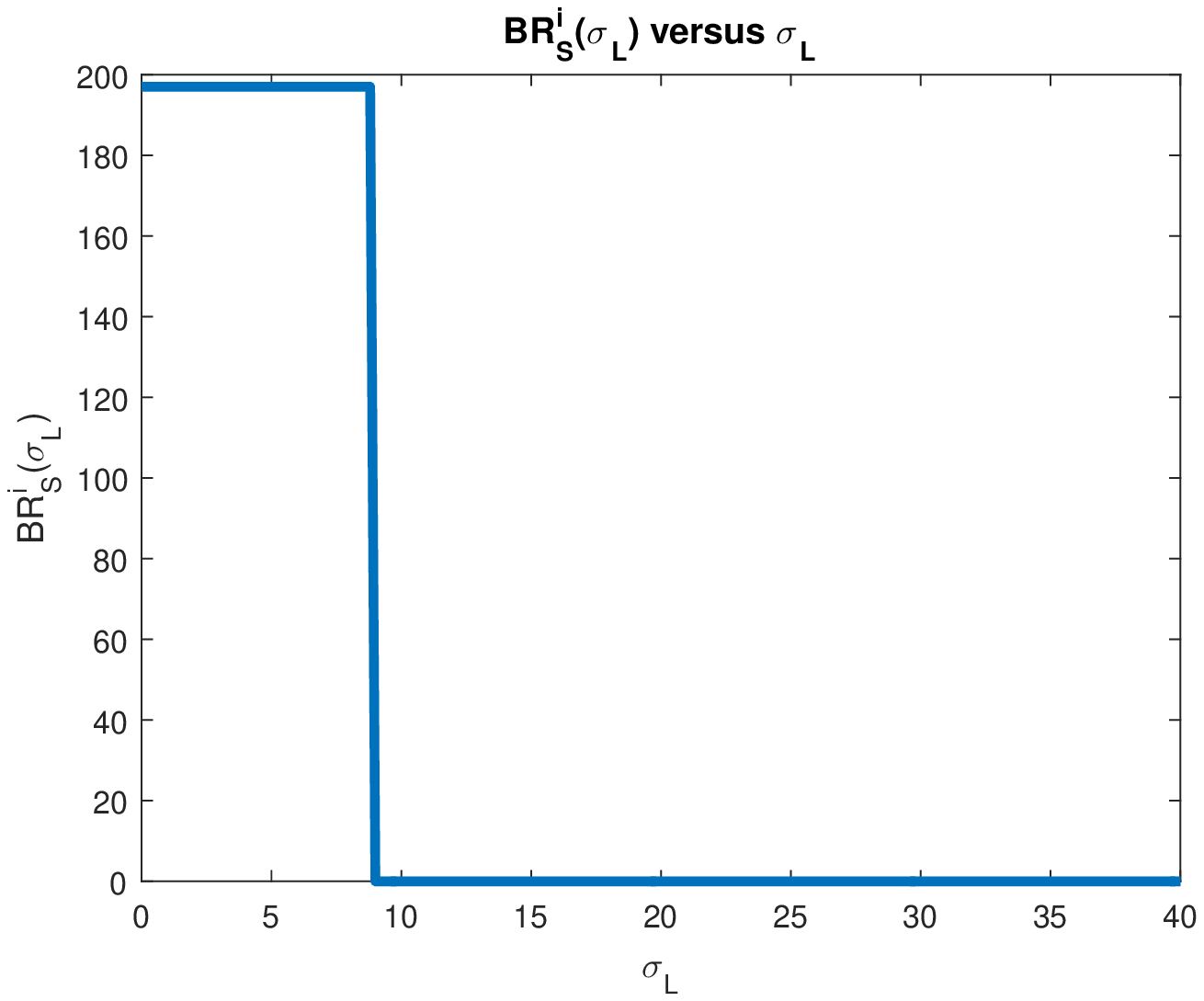}\includegraphics[width=0.25\textwidth]{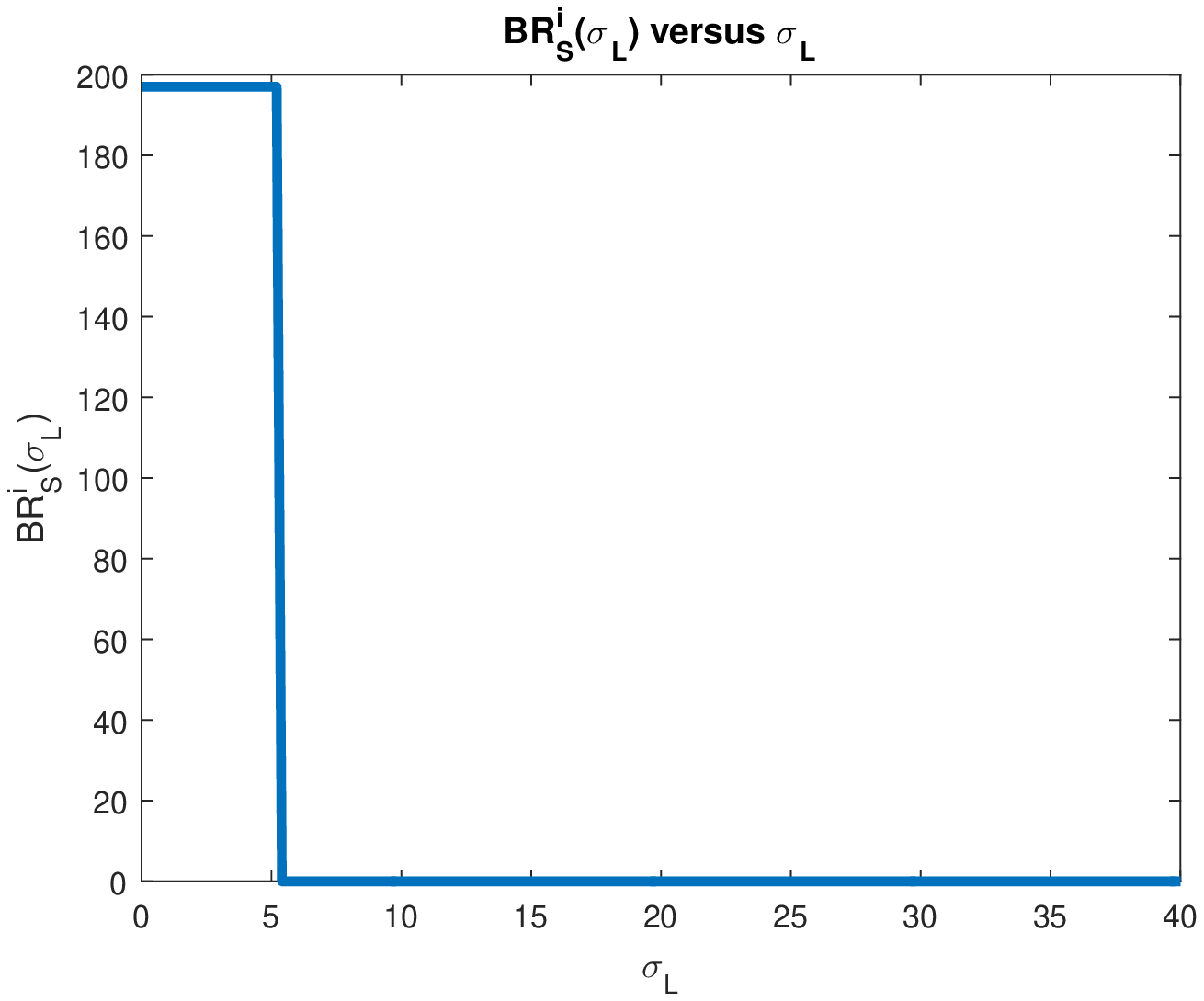}
\par\end{centering}

\begin{centering}
\includegraphics[width=0.25\textwidth]{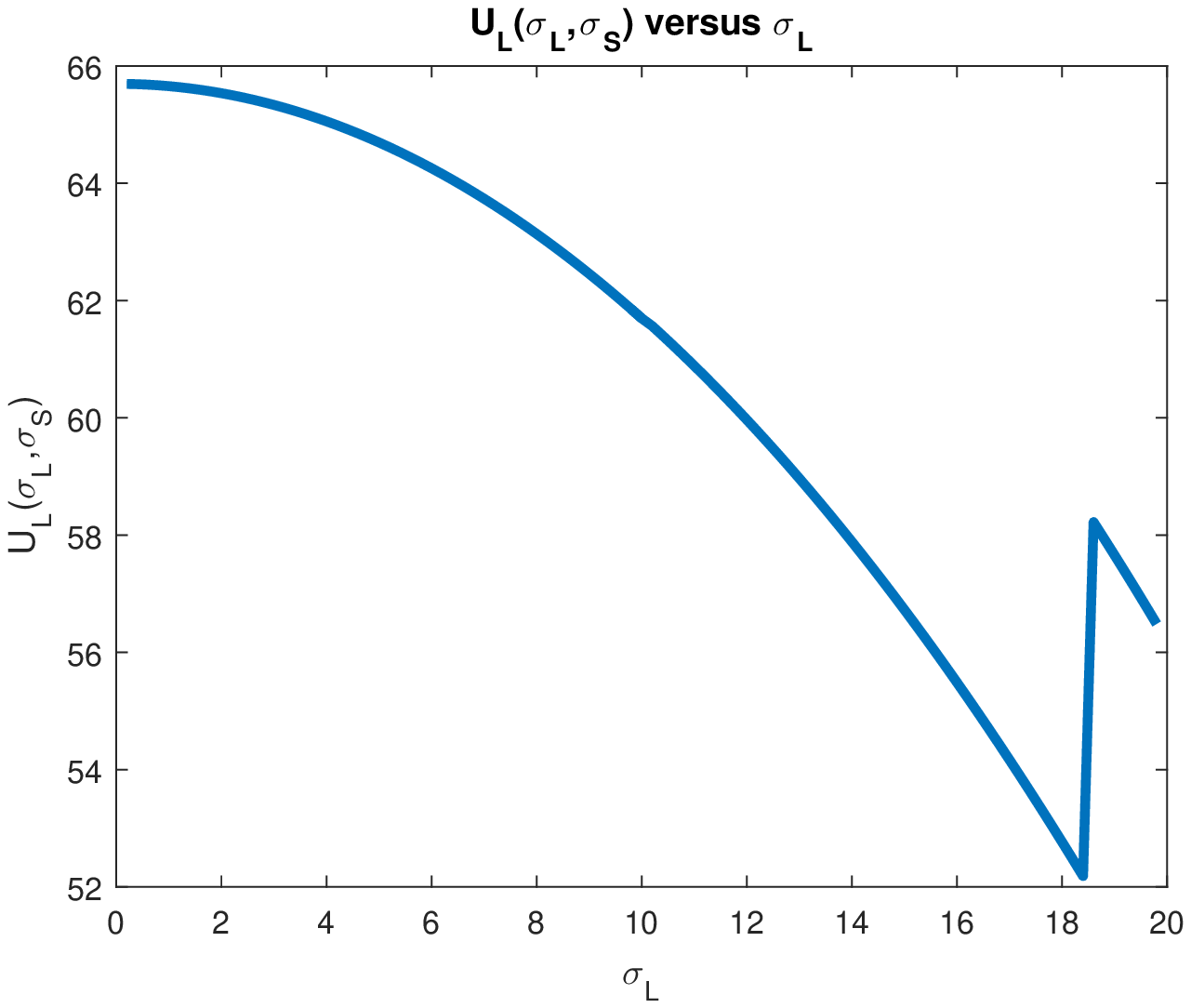}\includegraphics[width=0.25\textwidth]{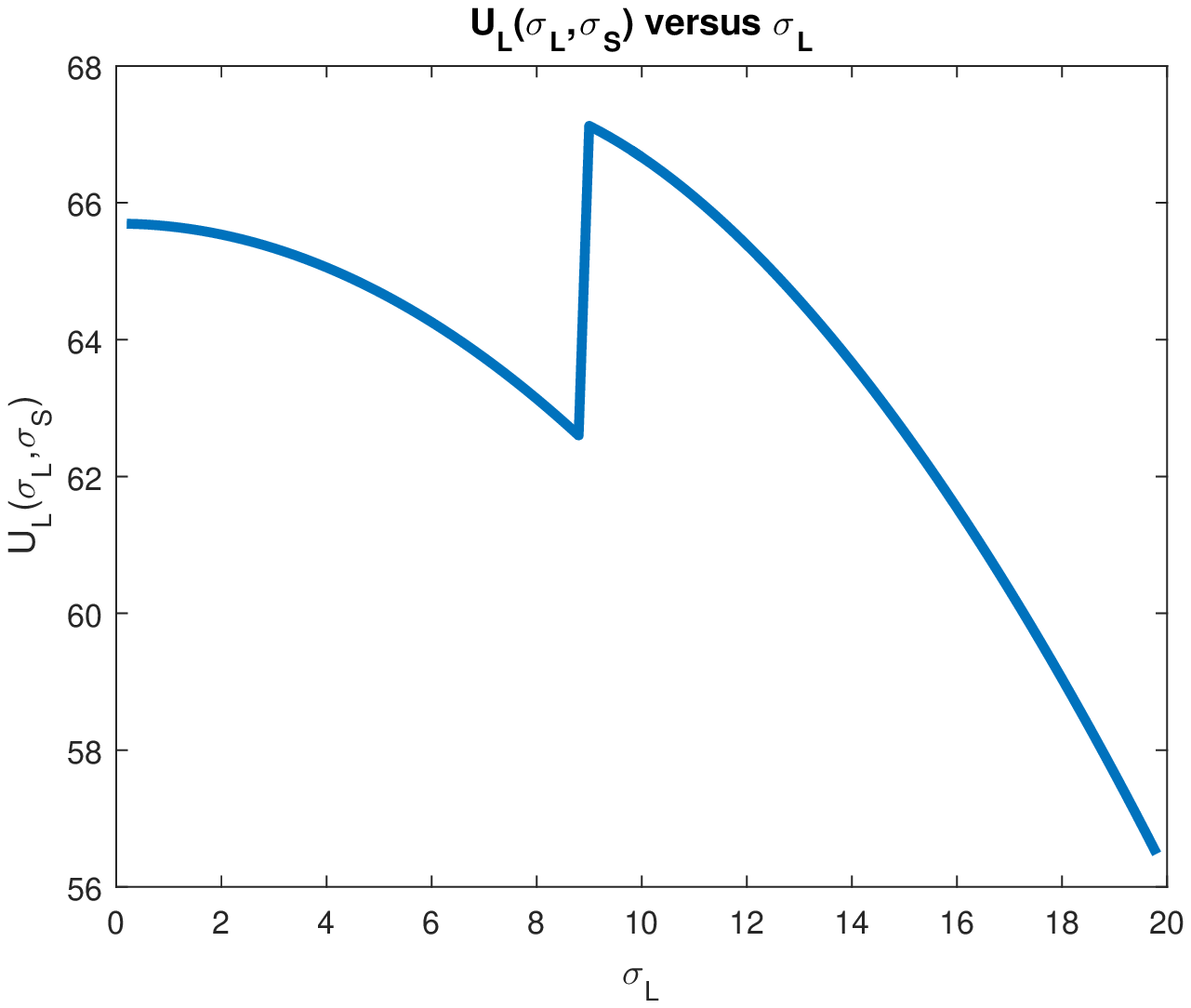}\includegraphics[width=0.25\textwidth]{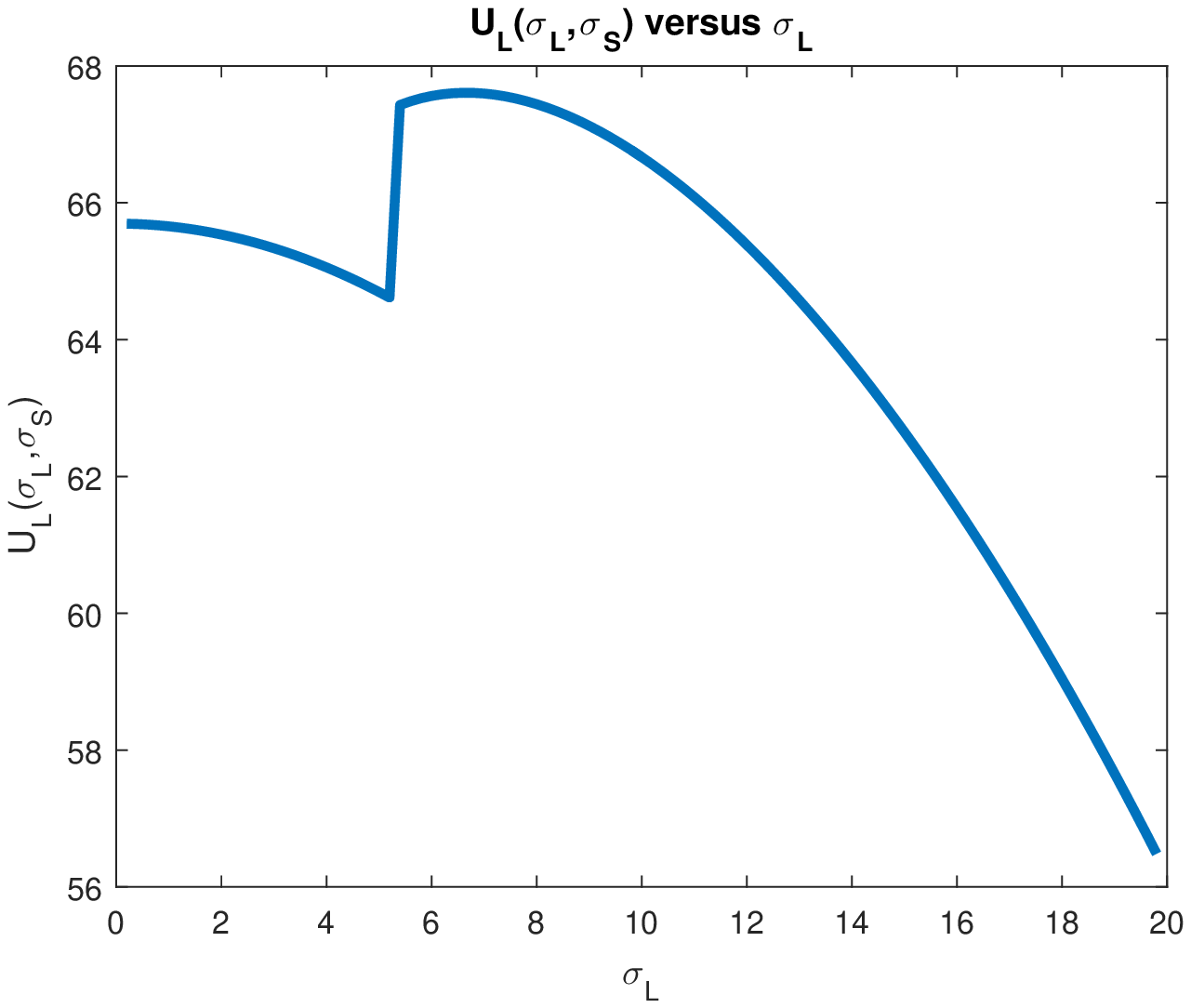}
\par\end{centering}

\caption{\label{fig:UtilSvPert}Row 1) User $i$ utility $U_{S}^{i}\left(\sigma_{L},\sigma_{S}^{-i},\sigma_{S}^{i}\right)$
versus $\sigma_{S}^{i}$ for various learner perturbation levels.
Row 2) $BR_{S}^{i}\left(\sigma_{L}\right),$ based on the $\sigma_{S}^{i}$
which achieves the highest utility for each curve in Row 1. Row 3)
Learner utility $U_{L}\left(\sigma_{L},BR_{S}^{i}\left(\sigma_{L}\right)\right),$
with $\sigma_{L}$ on the independent axis. From L to R, user perturbation
cost $\bar{N}_{i}^{S}=10,20,30.$ As $\bar{N}_{i}^{S}$ increases,
it becomes feasible for $L$ to perturb enough to discourage users
from perturbing.}
\end{figure*}
Because of the discontinuity in $U_{S}^{i}\left(\sigma_{L},\sigma_{S}^{-i},\sigma_{S}^{i}\right)$
introduced by the initial cost of perturbation, the best response
function $\sigma_{S}^{i*}=BR_{S}^{i}\left(\sigma_{L}^{*}\right)$
is cumbersome to solve analytically. Therefore, we solve for the Stackelberg
equilibrium numerically. Figure \ref{fig:UtilSvPert} displays the
results, in which the three columns represent user perturbation cost
$\bar{N}_{S}^{i}=10,20,30$ with other parameters held fixed.

Row 1 of the Fig. \ref{fig:UtilSvPert} depicts the optimization problem
of the users. For $\sigma_{S}^{i}>0,$ the users pick $\sigma_{S}^{i}$
which optimally balances their individual privacy-accuracy preferences.
This $\sigma_{S}^{i}$ could be large, because each user's perturbation
level affects his own accuracy only as one data point among many,
whereas it directly affects improves privacy. At exactly $\sigma_{S}^{i}=0,$
however, the user's utility jumps because he does not need to pay
the perturbation cost. Row 2 illustrates this bang-bang behavior,
which is summarized by Remark \ref{rem:BR}.
\begin{rem}
\label{rem:BR}At sufficiently-high $\sigma_{L}$ (the independent
variable), the users' privacy benefit becomes small enough that it
is outweighed by the cost of perturbation, and $BR_{S}^{i}\left(\sigma_{L}\right)$
falls to $0.$ As $\bar{N}_{S}^{i}$ increases (from left to right
in Fig. \ref{fig:UtilSvPert}), the $\sigma_{L}$ to dissuade user
perturbation decreases.
\end{rem}
This raises the question of whether the benefit of dissuading user
perturbation could be enough to justify the loss in accuracy and perturbation
cost of adding $\sigma_{L}.$ Remark \ref{rem:learner} states the
numerical result shown in Row 3 of the figure.
\begin{rem}
\label{rem:learner}In Column 1 ($\bar{N}_{S}^{i}=10$), the $\sigma_{L}$
required to dissuade user perturbation is sufficiently high so that
the benefits are outweighed by the loss in accuracy. In the other
columns, the accuracy loss that $L$ experiences due to her own perturbation
is overcome by the gain that she experiences when the users stop perturbing.
\end{rem}
In Columns 2 and 3, the jumps in $U_{L}$ are high enough that they
exceed the utility levels at $\sigma_{L}=0,$ and justify proactive
perturbation. In general, the higher the user perturbation cost $\bar{N}_{S}^{i},$
the less $L$ needs to perturb to dissuade users from perturbing.
The equilibrium in which $L$ perturbs proactively can be stated as
follows. 
\begin{enumerate}
\item Users prefer some privacy protection and are willing to invest in
technology for obfuscation if necessary. 
\item This obfuscation would be detrimental to $L.$ 
\item Instead, $L$ can perturb the data proactively.
\item $L$ need only match the users' desires for privacy \emph{up to their
perturbation costs} $\bar{N}_{S}^{i}.$ Then the users are satisfied
with $L$'s privacy protection and do not invest in obfuscation. 
\end{enumerate}

In some cases (\emph{i.e.}, Columns 2-4 of Fig. \ref{fig:UtilSvPert}),
$L$ improves his utility over cases in which the users perturb. Our
findings do not guarantee this result in all cases, but provide a
foundation for examining in which parameter regions $L$ can improve
his utility by protecting privacy proactively.

\section{Conclusion and Future Work\label{sec:Conclu} }

In this tracking-obfuscation interaction, the utility of each of the
users are interrelated, since they all affect the accuracy of the
output. Somewhat surprisingly, the optimal user perturbation levels
as functions of the learner perturbation level are independent of
one another. This leads to a self-interested behavior on the part
of the users and a high accuracy loss on the part of the learner.
In order to mitigate this problem, we have shown that a learner can
sometimes dissuade users from data obfuscation by proactively perturbing
collected information to some degree. Although she still must satisfy
the users' desired accuracy-privacy trade-off, she must only do so
to within some constant: the flat cost of user perturbation. If user
perturbation is sufficiently costly, privacy protection is incentive
compatible for the learner. For future work, we anticipate studying
an incomplete information version of the game, in which users' privacy
preferences are unknown, as well as a version of the game in which
the number of players is a random variable. These steps will help
to better understand and forecast the balance of power between user
obfuscation and machine learning.

\appendix

\section{Proofs of Accuracy Bound}

Theorem \ref{thm:accConstant} is proved using three lemmas. Lemma
\ref{lem:class} bounds the difference between the perturbed and unperturbed
classifiers. 
\begin{lem}
\label{lem:class}(Bound on difference between classifiers) Assume
that $\left|l'\left(z\right)\right|\leq1$ and $0\leq l''\left(z\right)\leq c.$
Then, for ERM with $L_{2}$-regularization, the magnitude of the difference
between the unperturbed classifier $\mathbf{f^{\dagger}}$ and the
input-perturbed classifier $\mathbf{f_{d}}$ is bounded in terms of
$\left\Vert \mathbf{f_{d}}\right\Vert $ by the deterministic quantity:
\begin{equation}
\left\Vert \mathbf{f^{\dagger}}-\mathbf{f_{d}}\right\Vert ^{2}\leq\frac{1+c^{2}\left\Vert \mathbf{f_{d}}\right\Vert ^{2}}{n^{2}\Lambda^{2}}\underset{i}{\sum}\left\Vert \mathbf{v_{i}}+\mathbf{w_{i}}\right\Vert ^{2}.
\end{equation}

\end{lem}
Essentially, the proof comes from comparing the first-order conditions
for each of the classifiers. Note that when norms are not specified,
we refer to the $L_{2}$-norm. Using this result, Lemma \ref{lem:emp}
bounds the difference in empirical loss.
\begin{lem}
\label{lem:emp}(Bound in difference in empirical loss) For any realized
database $D,$ the empirical loss is bounded by 
\begin{equation}
J\left(\mathbf{f_{d}},D\right)-J\left(\mathbf{f^{\dagger}},D\right)\leq\left\Vert \mathbf{f_{d}}-\mathbf{f^{\dagger}}\right\Vert ^{2}\left(1+c\right).\label{eq:CMempirical-1}
\end{equation}

\end{lem}
The proof of this lemma is based on work on empirical risk minimization
in \cite{chaudhuri2011differentially}. The next step is to bound
the difference in expected loss using the difference in empirical
loss. The result is given in Lemma \ref{lem:exp}.
\begin{lem}
\label{lem:exp}(Bound in difference in expected loss) The difference
in expected loss due to $\mathbf{f_{d}}$ and $\mathbf{f^{*}}$ satisfies,
with probability $1-\delta,$ 
\begin{equation}
\hat{J}\left(\mathbf{f_{d}}\right)-\hat{J}\left(\mathbf{f}^{*}\right)\leq2\left[J\left(\mathbf{f_{d}},D\right)-J\left(\mathbf{f^{\dagger}},D\right)\right]+O\left(\frac{\text{log}\left(1/\delta\right)}{\Lambda n}\right).\label{eq:expLossInter}
\end{equation}
Define $\mathbf{u_{i}}\triangleq\mathbf{v_{i}}+\mathbf{w_{i}}.$ Using
Lemma \ref{lem:class} and Lemma \ref{lem:emp}, with probability
$1-\delta,$ $\hat{J}\left(\mathbf{f_{d}}\right)-\hat{J}\left(\mathbf{f}^{*}\right)\leq$
\begin{equation}
\frac{2+2c^{2}\left\Vert \mathbf{f_{d}}\right\Vert ^{2}}{n^{2}\Lambda^{2}}\underset{i}{\sum}\left\Vert \mathbf{u_{i}}\right\Vert ^{2}\left(1+c\right)+O\left(\frac{\text{log}\left(1/\delta\right)}{\Lambda n}\right),\label{eq:expLossBound}
\end{equation}

\end{lem}
Equation \ref{eq:expLossInter} is from Theorem 1 of \cite{sridharan2009fast},
which bounds the difference between the expected loss of any classifier
and the optimal classifier. Next, we bound $\left\Vert \mathbf{u_{i}}\right\Vert ^{2}$
with some probability.
\begin{lem}
\label{lem:probBound}(Bound on error realization of random variables)
Since $\mathbf{v_{i}}\sim\mathcal{V}_{i}$ and $\mathbf{w_{i}}\overset{\text{i.i.d.}}{\sim}\mathcal{W},$
$\mathbf{u_{i}}$ are draws from the distribution $\mathcal{V}_{i}+\mathcal{W}.$
From the cumulative distribution function of the $\chi^{2}$ variable,
the square of their magnitude can be bounded with some probability
by 
\begin{equation}
\mathbb{P}\left\{ \left\Vert \mathcal{\bar{V}}_{i}+\mathcal{\bar{W}}\right\Vert ^{2}\leq\zeta\left(\sigma_{L}^{2}+\left(\sigma_{S}^{i}\right)^{2}\right)\right\} =\frac{\gamma\left(\frac{d}{2},\frac{\zeta}{2}\right)}{\Gamma\left(\frac{d}{2}\right)}.\label{eq:realizationRV}
\end{equation}

\end{lem}
The probability that the bound in Eq. \ref{eq:realizationRV} fails
and that the bound in \ref{eq:expLossBound} fails is the product
of the probability that each individually fails. Thus a conservative
bound is $\hat{J}\left(\mathbf{f_{d}}\right)-\hat{J}\left(\mathbf{f}^{*}\right)\leq$
\begin{equation}
\frac{2+2c^{2}\left\Vert \mathbf{f_{d}}\right\Vert ^{2}}{n^{2}\Lambda^{2}}\underset{i}{\sum}\zeta\left(\sigma_{L}^{2}+\left(\sigma_{S}^{i}\right)^{2}\right)\left(1+c\right)+O\left(\frac{\text{log}\left(1/\delta\right)}{\Lambda n}\right),\label{eq:finalAccBound}
\end{equation}
with probability at least $1-\delta\left(1-\gamma\left(\frac{d}{2},\frac{\zeta}{2}\right)/\Gamma\left(\frac{d}{2}\right)\right).$
This result leads to Theorem \ref{thm:accConstant}.\bibliographystyle{plain}
\bibliography{WIFSbib}

\end{document}